# Improving Statistical Multimedia Information Retrieval (MIR) Model by using Ontology

Gagandeep Singh Narula
B.Tech, Guru Tegh Bahadur Institute of Technology, GGS Indraprastha University, Delhi

Vishal Jain
Research Scholar, Computer Science and Engineering Department, Lingaya's University, Faridabad

## ABSTRACT
The process of retrieval of relevant information from massive collection of documents, either multimedia or text documents is still a cumbersome task. Multimedia documents include various elements of different data types including visible and audible data types (text, images and video documents), structural elements as well as interactive elements. In this paper, we have proposed a statistical high level multimedia IR model that is unaware of the shortcomings caused by classical statistical model. It involves use of ontology and different statistical IR approaches (Extended Boolean Approach, Bayesian Network Model etc) for representation of extracted text-image terms or phrases.

A typical IR system that delivers and stores information is affected by problem of matching between user query and available content on web. Use of Ontology represents the extracted terms in form of network graph consisting of nodes, edges, index terms etc. The above mentioned IR approaches provide relevance thus satisfying user's query.

The paper also emphasis on analyzing multimedia documents and performs calculation for extracted terms using different statistical formulas. The proposed model developed reduces semantic gap and satisfies user needs efficiently.

## State of Art
Research on multimedia information retrieval seems to be gargantuan and challenging task. Its areas are so diversified that it has lead to independent research in its own components. Firstly, there used to be human centered systems that focus on user's behavior and needs. Various experiments and studies were conducted in lieu of these systems. The users were asked to present a set of valuable things in daily life. It was done on similarity of users. Some of choices are same while some are different. Few of them prefer to use images instead of text caption.

In further experiments, it was noticed that new users were taking feedback from previous users. It leads to concept of relevance feedback module in information model. In early years, most research was done on content- based image retrieval. The existing models are of different level and scope. These models are semantically unambiguous. For e.g.: IPTC model **[1]** uses location fields that focus on location of data but this model also failed due to lack of statistical approach. Another metadata model was developed i.e. EXIF **[2]** to support features of images but it did not tell anything about relationship and associations between different contents of image. It also resulted in vain. The third model developed was Dublin Core **[3]** that deals with semantic as well as structural content of image and text but it failed to depict relationship between text and image.

With advancement in technology and predictions, some probabilistic and futuristic models were also developed. In following paper, statistical multimedia IR model has been proposed and compared with classical multimedia IR model.

## 1. INTRODUCTION
Human knowledge is richest multimedia storage system. There are various mechanisms like vision, language that expresses knowledge and information obtained from them must be processed by system efficiently. There must be systems designed that interprets and process human queries, thus producing relevant results. It is often seen that users get baffled while searching results of their queries. The reasons behind this are:

- The content of information is unclear and needs user to refine that information.
- The data stored on systems may or may not be updated regularly.
- There lies lower level of interaction between user request and stored information on systems. The low-level links are called Semantic Gap.

Statistical approaches involves retrieved documents that matches query closely in terms of statistics i.e. it must have statistical model, calculations and analysis. These approaches break given query into TERMS. Terms are words that occur in collection of documents and are extracted automatically. For reducing inconsistencies and semantic gap in multimedia information, it is necessary to remove different forms of same word because it makes user confused in choosing specific terms that lies close to query. Some IR systems extract phrases from documents. A phrase is a combination of two or more words that is found in document.

We have used approaches like extended Boolean approach, network model that performs structural analysis for retrieving text or image pairs. They also assign weights to given term. The weight is defined as measure of effectiveness of given term in distinguishing one document from other documents. The paper has following sections: Section 2 describes architecture of classical multimedia model. Section 3 lets reader go through proposed IR model that is implemented using statistical approaches with the use of ontology. It also requires conversion of low level features to high level features using multimedia analysis. Section 4 deals with experimental analysis and calculations depicting the relevance of proposed model. Finally, Section 5 concludes about paper.

## 2. CONCEPT OF MULTIMEDIA IR SYSTEM

The classical multimedia IR system has not proven effective in extraction of relevant terms from document collections. Traditional IR systems are not intelligent that they are able to produce accurate results. These systems use human perception to process query and returns results. The results may be relevant or non- relevant because these systems match query with information stored in information database.

The syntax of multimedia document is different from text documents. Multimedia documents do not contain any information symbols or keywords that help in expressing information. They consist of:

- Visible and Audible Data Types: - It includes text, images, graphs, videos and audio.

- Structural Elements: - They are not visible. They describe the organization of other data types.

The salient features of multimedia information **[4]** are given below:

- The information stored in document that is to be searched can be audio, visual, videos etc. They communicate variety of messages and emotions that helps to understand easily.

- Structure information gives organization and usability in performing communications.

- There is communicational gap between user and system. It is known that some systems are fast in processing of calculations whereas human is not. So, it leads to communication gap.

### 2.1 Layout of Classical Multimedia Ir Model

Since multimedia documents do not contain keywords or symbols that facilitates easy process of searching through document. Keeping this in mind, this classical model consists of Query Processing Module that translates the multimedia information tokens into symbols / keywords which are easily understood by system. The model has following modules:

- Analysis Module: - IR system firstly analysis multimedia documents and extract features from them. The features include low- level as well as high- level features.

- Indexing Module: - The module that stores features or terms retrieved from multimedia documents is called Indexing Module.

- Query Processing Module: - This module translates multimedia information tokens like audio, text-pairs, videos etc into information symbols that are now understood by system.

- Retrieval Module: - It finds rank of stored documents on basis of similar terms used in query. After ranking of documents, the results satisfying query are presented to user.

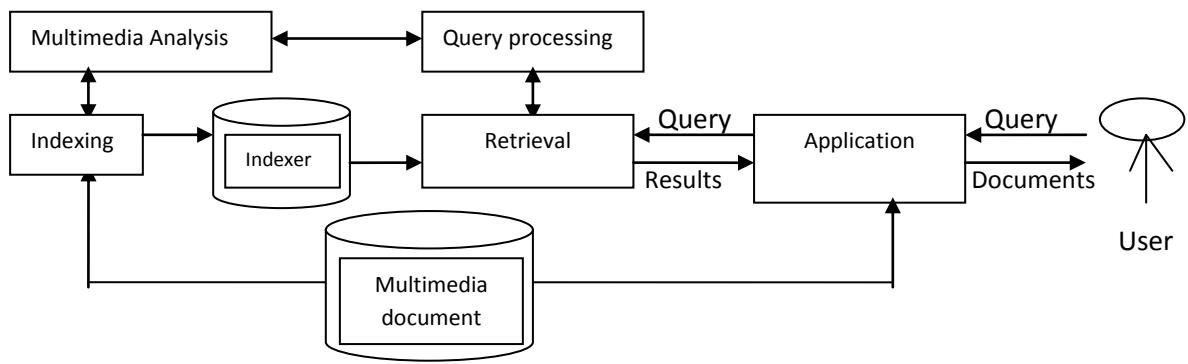

**Figure1: A Classical Multimedia IR Model [5]**

### 2.2 Shortcomings of Classical Multimedia IR Model

They are explained below:

- The classical model deals with terms or information symbols instead of maintaining relationships between them. It does not give any information about concepts used in extracted terms or image pairs.

- It creates semantic gap **[6]** between user and system due to availability of irrelevant and superfluous information terms stored in information database of IR system.

- It does not involve concept of ontology and semantic associations for representing concepts associated with terms in document.

- The terms which are relevant and similar to each other are identified at the end of phase by RETRIEVAL Module. The good model is one that has capability to distinguish between relevant and non relevant terms in the middle of phase in order to prevent any confusion.

- The model does not involve the concept of re-use of queries. Once the query is expanded, it will not store in system for future use. Again, it has to analyze large collection of documents and retrieve terms from them.

- It does not employ any statistical or probabilistic approaches for determining relevance of IR system.

## 3. PROPOSED HIGH LEVEL MULTIMEDIA IR MODEL

A model is being designed that employ use of statistical IR approaches for extracting terms from multimedia documents. Ontology Module has been introduced that serves the task of representing concepts and relationships among retrieved terms. In order to overcome this problem, the model includes only those approaches that perform extraction of terms like images, video, and text from multimedia documents as well as text documents.

The block diagram of proposed model containing several modules is shown below:

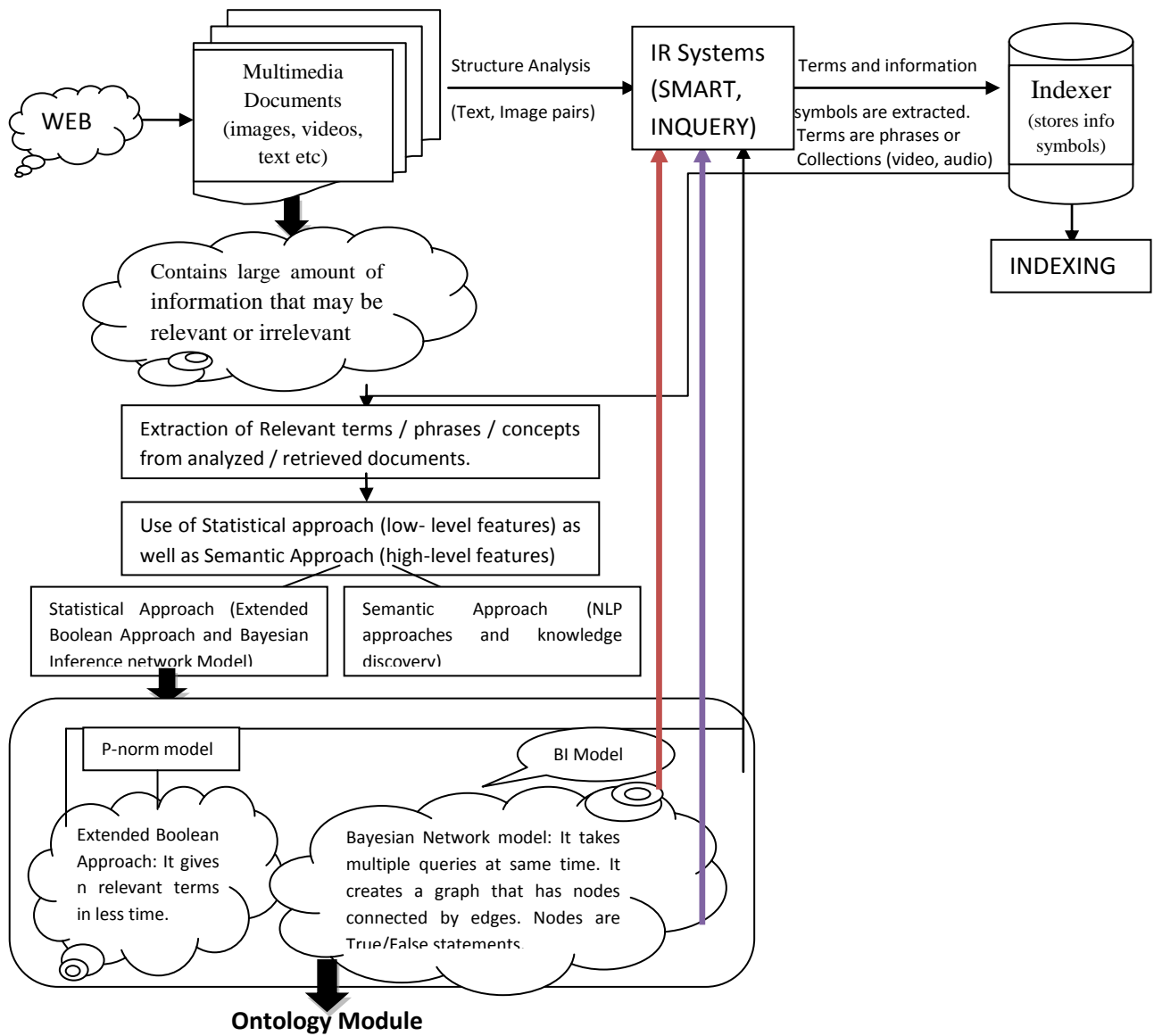

**Figure2 (a): Proposed High- level Statistical Multimedia IR Model**

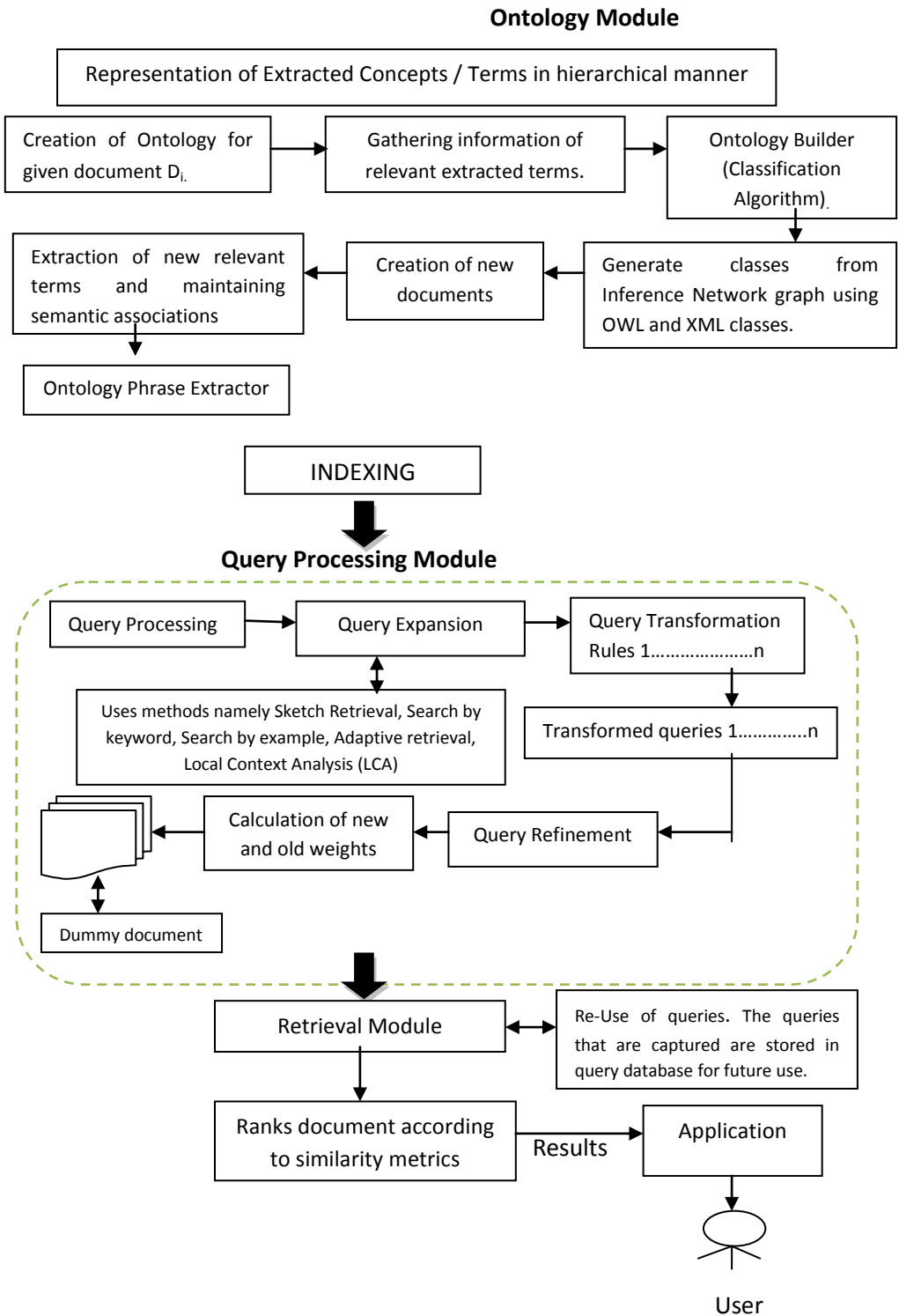

**Figure2 (b): Proposed High- level Statistical Multimedia IR Model**

The model has following aspects:

- Improves user expressiveness: - It analyses terms that have close meaning to user's query and expressive results are presented to user.

- Supports different modules: - Several modules like Ontology Module, Extraction Module, Query Expansion and Refinement have been introduced in proposed model.

- Low Computation and Cost: - The approaches that are used to extract terms from documents are so efficient that they takes into account only relevant terms and discards non relevant terms. Only relevant terms are expanded and it leads to saving of time and work.

- Good Retrieval Accuracy: - The model retrieves only those terms from documents that satisfies user's information needs.
- Pipelining Facility: - Pipelining means dividing of complex tasks into certain number of independent sub tasks that performs parallel to each other. It helps in extraction of text- image documents by dividing into smaller segments. Each segment holds some information. As soon as each part is analyzed, terms from different segments are retrieved and combined to produce full document.

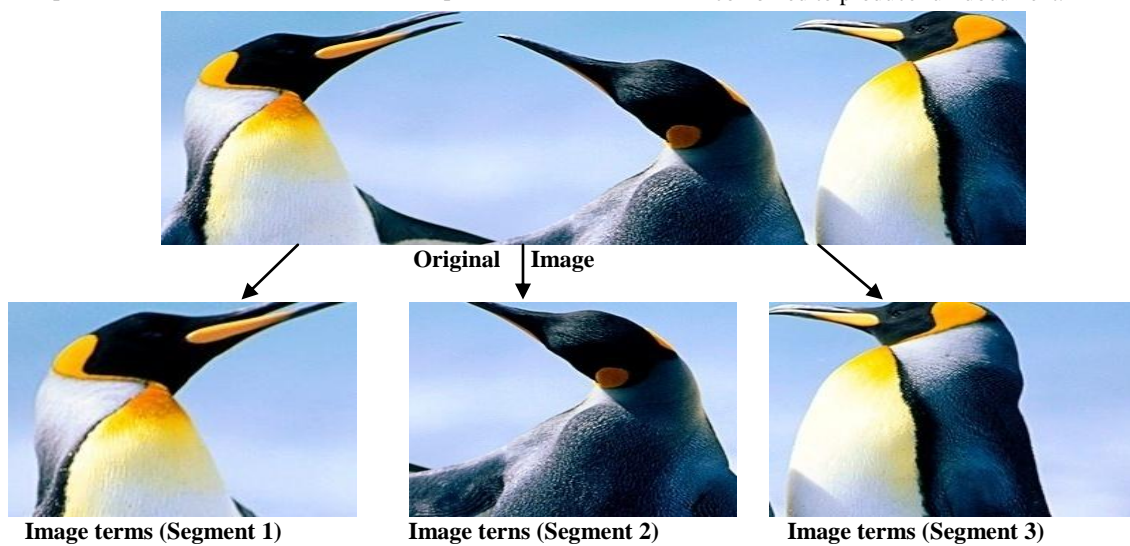

**Original Image**

**Image terms (Segment 1)**     **Image terns (Segment 2)**     **Image terms (Segment 3)**

**Figure3: Extraction of Image Terms [7]**

## 3.1 Multimedia Document Analysis Module
There is large number of multimedia documents consisting of video text collections on web. The IR systems used in model performs structural analysis of documents and extracts text- image terms from them **[8]**. At this stage, it is not possible to fully determine that random chosen documents are relevant or non relevant. The classical model works on low level multimedia analysis. The proposed multimedia model works on high level multimedia analysis algorithm rather than low- level analysis because of following reasons:

**Table1: Features of High – Level Multimedia Analysis**

| Low- Level Multimedia Analysis | High- Level Multimedia Analysis |
| --- | --- |
| 1. It produces low level features like text and image terms. | 1. It produces high level features like it describes concepts associated with extracted low level text terms. |
| 2. It only extracts relevant terms from information that is stored in information database on system. | 2. It extracts terms from derived documents even if information is not stored on system. |
| 3. It uses information symbols to build the index of multimedia documents. These symbols may or may not specify given concepts. | 3. It uses keywords that are related to document and always show presence of concepts that are described by terms in given document. |

## 3.2 Indexing Module
The terms and information symbols are extracted in previous module. The storage location of these extracted terms (relevant/ irrelevant) is decided by Indexing Module. It is done with the help of Indexer that stores the generated terms. This module has capability to store high dimensional information i.e. it can also solve structured indexes or trees along with information symbols.

## 3.3 Extraction Module
This module uses two or more statistical approaches for extraction of relevant terms or phrases from retrieved documents. It is a module that determines the relevance of IR system. It is able to provide distinction between relevant and non relevant terms on basis of results produced by statistical approaches. The approached are discussed in following sub sections:

### 3.3.1 Extended Boolean Approach
*Problem*: - The classical Boolean condition i.e. True/ False produces both relevant and non relevant results. They supply given solution in response to whole document. If some text or image terms are relevant in document and some are not relevant, then Boolean condition leads to irrelevant results because it considers whole document.

*Solution*: - Extended Boolean Approach

*Analysis*: - A number of extended Boolean models have been developed to provide ranked output of results i.e. the documents that satisfies user's query. These models use extended Boolean operators that are called as Soft Boolean Operators for finding relevant text- image pairs. This approach assigned different weights to different terms and computes relevance.

The classical Boolean operators are different from Soft operators as follows:

**Table2: Classical Operators vs. Soft Operators**

| Classical Boolean Operator | Extended Operators (Soft Operators) |
|---|---|
| It evaluates its terms to return two values only i.e. True/False. The values are represented by zero (False) and 1 (True) respectively. It is represented by truth tables graphically. | It evaluates its items to number on basis of degree to which condition matches document i.e. If condition matches document, then it returns 1 else 0. If some part satisfies condition while other part does not, then the value is in fraction. It means soft operators do not leave document as irrelevant. |

Example of Extended Boolean approach is p-norm model.

*P-Norm Model*: - The model performs evaluation if and only if terms satisfy user's query in accordance with user's views. The model uses two functions AND, OR for finding similar documents and terms. Consider a query that has n terms given by $q_1, q_2, q_3 \ldots q_{n-1}, q_n$ with corresponding weights $wq_1, wq_2, wq_3 \ldots wq_{n-1}, wq_n$ in a given document $D_i$. The document is also assigned weights as $wd_1, wd_2, wd_3 \ldots wd_{n-1}, wd_n$.

Firstly, the extended Boolean function AND finds similar documents by combining (AND) query terms together. Then, terms are retrieved from those documents that satisfy user needs. AND function follows condition that all components must be present in order to return relevant (non zero) terms. If any component is absent, then it will give zero values.

(1) $S_{AND}$ (d $(q_1, wq_1)$ AND ............ AND $(q_n, wq_n)$) = $1 - [(\sum(1-wd_i)^p * (wq_1)^p) / (\sum(w_q)^p)]^{1/p}$

Where $1 \leq p \leq \infty$ and $S_{AND}$ = Similar documents retrieved using AND function.

The Extended Boolean OR function finds similar documents with query that add (OR) the query terms together.

(2) $S_{OR}$ (d $(q_1, wq_1)$ OR ............ OR $(q_n, wq_n)$) = $[(\sum(wd_i)^p * (wq_1)^p) / (\sum(w_q)^p)]^{1/p}$

Where $1 \leq p \leq \infty$ and $S_{OR}$ = Similar documents retrieved using OR function

So, we conclude that p- norm model returns n relevant multimedia terms instead of binary terms. It reduces system time and increases performance.

*Drawbacks*: Extended Boolean approach fails in extracting relevant terms from given n terms. P- norm model assigns weights to query terms as well as document terms. Both queries are treated equally because p – norm functions evaluate all term weights in a same way. It cannot distinguish between relevant and non- relevant terms. The solution to this problem lies in usage of probabilistic statistical IR approaches.

### 3.3.2 Bayesian Probability Models / Conditional Probability Models

Bayesian models give relationship between probability of random selected documents and probability that given document is relevant. In such case, we are aware of features of document (image terms, text, statistics, phrases etc) and then calculate its probability. Following are features of probabilistic models: -

- They are related to prior and posterior probabilities. Prior means finding probability as earliest as possible without knowing features of document. Posterior means finding probability after examining the features of document. **Prior Probability + Posterior Probability = 1**

- Conditional Models are also called as Probability Kinematics model that is defined as flow of probabilities of relevant terms to non relevant terms in whole document.

- It uses concept of Inverse Document Frequency (idf) for determining number of relevant terms by using formula as:

  Idf = ln N/n where N= No of total documents, n = No of relevant documents

- Probabilistic models helps in achieving relevance on basis of values estimated for different documents.

The statistical probabilistic models **[9]** are categorized into two parts:

(a) **Binary Independence Model (BI)**: - The model in which each text- image term (relevant/ irrelevant) is independent of other text- image pairs in collection of documents is called BI model. So, the probability of any relevant/ irrelevant term is independent of probability of any other terms in documents.

BI model is also called *Relevance Weighting Theory*. It says that each term is given weight that is used to rank documents on basis of relevance, thus extracting relevant terms. *Weights are assigned by product of Term Frequency and Inverse Term Frequency i.e. (tf * idf) when we are taking random collection of documents*. Term Frequency (tf) means number of terms occurred in document. So, tf varies from one document to another whereas Inverse Document Frequency (idf) measures how many times the given term occurs in document. It gives probability of terms occurred in a document.

Consider number of finite terms $t_k$ in document $d_i$. Each term is assigned different weights $w_k$ that is to be calculated according to given formula:

$W_k = \log [P_k (1- U_k) / U_k (1 – P_k)]$ (When we are given set of data terms)

Where $P_k$ = Probability of term $t_k$ occurring in relevant documents

$U_k$ = Probability of term $t_k$ occurring in non relevant documents

$W_k$ = Weight to each term. It is defined as measure of distinguishing relevant terms from non relevant terms. It is also called as Term Relevance Weight or Log Odds Function.

Odd ratio is calculated on basis of likelihood of terms in relevant documents as well as in non relevant documents. Let likelihood of terms in relevant documents is $X = (P_k / 1- P_k)$ and in non relevant documents $Y = (U_k / 1-U_k)$. Then $W_k$ is given by X / Y. $W_k$ is zero if $P_k = U_K$, $W_k > 0$ if $P_k > U_k$

The model concludes that the terms which occur many times in single document is relevant but if same terms occur in large number of large number of documents, then it is not relevant. So, a weight function is developed that varies from idf to $W_k$ formula.

*Limitation of this model*: - It is not able to distinguish between low frequency terms and high frequency terms in context of weights. It gives weight of low frequency terms as same as those of high frequency terms. It does not able to extract terms from multiple queries also. So, to overcome these problems, we have used Inference Network Model.

(b) **Bayesian Inference Network Model**

It is one of statistical approach for extraction of terms from multimedia documents with the help of constructing graph called as Inference Network Graph. Besides computing probabilities for different nodes, this model also determines concepts between various retrieved terms. It provides surety that user needs are fulfilled because it also combines multiple sources of evidence regarding relevance of document to user query.

*Graph Structure*: - Inference Network is a graph that has nodes connected by edges. Nodes represent True/ false statements describing term is relevant or not. A graph has following elements:

- Document Nodes ($D_n$) : - They are called Root Nodes

- Text Nodes ($T_n$): - They are child nodes of document nodes. It may include audio, video nodes, text image nodes etc. So, child nodes have multiple representations of document.

- Concept Representation Nodes ($CR_n$): - They are child of text nodes. The concepts used in terms that are in text nodes are represented by CR nodes. These nodes are index terms or keywords that are matched in document and retrieves relevant terms.

- Document Network: - It is network consisting of Document nodes, Text nodes, and CR nodes. It is not tree as it has multiple roots and nodes. Document Network is Directed Acyclic Graph (DAG) since it has no loop. The representation of document network for different documents from $D_1$ to $D_n$ is shown as:

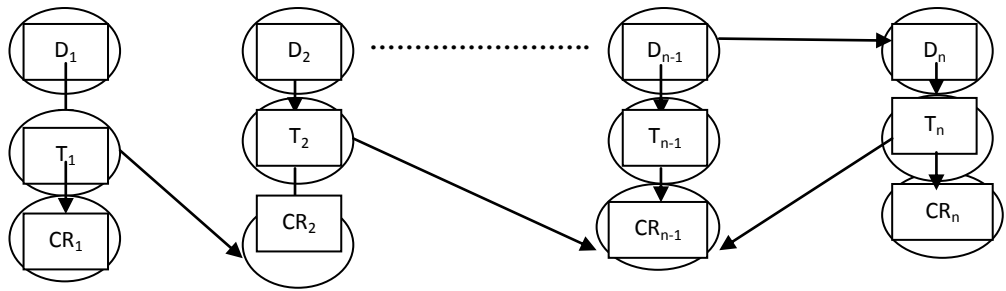

**Figure4: Document Network (It describes concepts used in multiple terms from different documents)**

- Query Network: - Since we have extracted concepts in Document Network, it is possible that different concepts are used in same query nodes or different concepts in different nodes. The concepts that describe relevant terms are shown in form of results and presented to user.

The representation of query network for different query nodes from $Q_1$ to $Q_n$ is shown as:

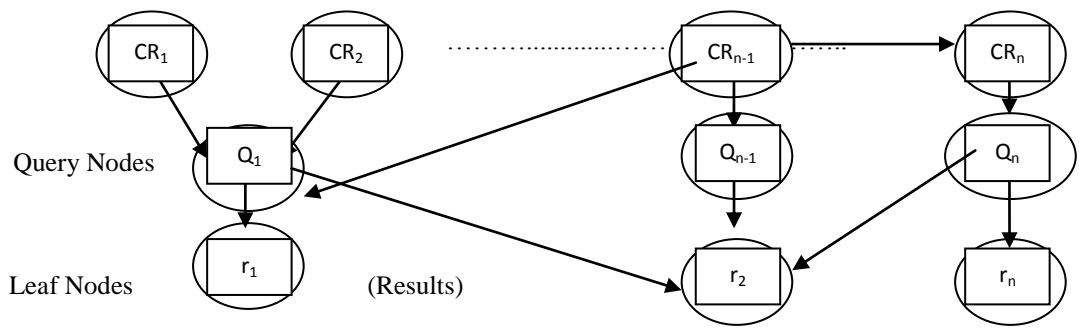

**Figure5: Query Network (It describes generation of results (leaf nodes))**

When we combine document network and query network, we get inference graph. This graph computes probabilities of terms contained in child nodes of document nodes and so on. It is done by using LINK MATRIX. Each node is assigned with its weight in each row of matrix. The column represents number of possible combinations a node can have.

In link matrix, *Number of parents = $2^n$ Number of columns*. If node has 3 parents, then there will be 8 columns. Then, probabilities and weight function are computed for all 8 columns of matrix. Each probability is multiplied by its weight and then all eight probabilities are added to get total probability of their respective parents' node. Consider combination of 110 (1 stands for True, 0 for False). The probability for combination is calculated as P1 * P2 * (1-P3). Weight function for such combination is (W1 + W2) / (W1 + W2 + W3). Total probability is calculated as P1 * P2 * (1-P3) * (W1 + W2) / (W1 + W2 + W3).

## 3.4 Ontology Module

This module is used to represent concepts and conceptual relationships among nodes that are described by inference network graph in previous module using concept of ontology. Ontology is defined as Formal, Explicit, and Shared

Conceptualization of concepts, thus organizing them in hierarchical fashion **[10]**. Various phases of ontology module are described below: -

(a) *Creation of Ontology or Ontology Representation*: - Inference Graph consists of document nodes ( root nodes).

$D_i$ has concept nodes as $CR_i$.
Edges represent relationship between them

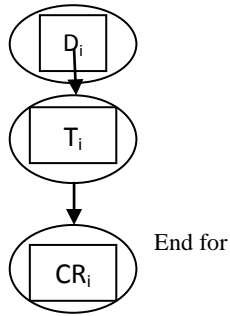

Each document node has concept nodes that are treated as Vertices. An edge from one node to other node represents relationship among concepts.

(b) *Ontology Building*: - It uses an algorithm for developing ontology for inference graph. It requires use of OWL (Ontology Web Language) that is used for writing ontology. It is used for creating objects of each class.

 BEGIN

For each vertices V of inference graph G

Class C = new (owl: class)

C.Id = C.label                              // each concept has its unique identification and name//

DatatypeProperty DP = new (owl: DatatypeProperty)   // DatatypeProperty require values with that, should be type of

DP.Id = DP.Name, DP.Value;

DP.AddDomain (C);                           // It adds values of child nodes to given concept node C//

For each edge E of Graph G

DP.AddDomain (B.getClass ())                // getClass is used to show relationship between concepts//

End for

End begin

(c) *Generation of OWL class*

Class Result = new (owl: class)             // Result represents leaf nodes//

Result.Id = Result. Name

DatatypeProperty ResultDP = new (owl: DatatypeProperty)
// to show value of leaf nodes//

ResultDP.Id = Result.Name, Result. Value;
// Leaf nodes have name and value//

Result.AddDomain (Result)

For each edge E of Graph G

Class Relationship = new (owl: class)

Relationship.Id= "       "

For each vertices of graph

Relationship.Id= Relationship.Id + C.label;

End for

ResultDP.AddDomain (Relationship)

End for

### 3.5 Query Processing Module
A query is called information need. It is final result with optimal and effective terms. This module deals with expansion and refinement of query either automatically or manually with user interaction. It analyze query according to query language, extract information symbols from it and pass it to Retrieval Module for searching index terms.

*Query Expansion through manual methods*: It includes:

- Sketch Retrieval: - It is one of methods to query a entity of required database. With that, query is visual sketch given by user, and then system processes this drawing to extract its features and searches the index for similar images.

- Search by Example: - In this, user gives query as an example of image that he tends to find. A query then extracts low level features.

- Search by Keyword: - It is most popular method. User describes information with set of relevant terms and system searches it in documents.

*Query Expansion through automatic method*: - It includes Local Context Analysis (LCA) approach.

It is one of best methods for automatic query expansion. It expands terms from query, rank and weights them by using certain formula.

**LCA = Local Feedback Analysis + Global Analysis**

It is local because concept relevant terms are only retrieved from globally retrieved documents. It is global because documents related to given query topic are selected randomly from huge collection of documents present on web (like we have selected three documents related to semantic web from web). When we put query in Google and press ENTER, query is executed and it retrieves some documents. It is global activity. LCA is concept based fixed length scheme. It expands user query and retrieves top n relevant terms that closely satisfies query. It returns only fixed number of terms.

The retrieved terms are ranked accordingly as:

**Belief (Q, C) = $\Pi$ [$\varphi$ + log (af(c, $t_a$)) idfc / log (n)] $^{idfa}$**

Where C= Concepts related to query Q

Belief (Q, C) = Ranking Function

$t_a$ = Query Term

$af(c, t_a) = ft_{a1d1} * f_{c1d1} + ft_{a2d2} * f_{c2d2} + ft_{a3d3} * f_{c3d3} + \ldots\ldots\ldots ft_{an, dn} * f_{cn, dn}$

$$af(c, t_a) = \sum_{d=1}^{d=n} ft_{ad} f_{cd}$$

Where d = documents from 1 to n

$ft_{ad}$ = Frequency of occurrence of query term $t_a$ in document d

$f_{cd}$ = Frequency of concepts (terms related to query) in document d

idfc = It measures importance of concepts related to query terms i.e. how many times the same concept is used in document

idfa = It measures importance of query terms $t_a$.

φ = It is constant used for distinguishing between relevant and non relevant terms. It stores non relevant terms that are treated as constant.

### 3.5.1 Query Refinement

A tern can have different weights in each relevant document, so there is need to refine query. Query Refinement means calculation of old weights of expanded query terns in order to produce new weights of same query terns. These query terms are transformed into dummy document that is used for Indexing.

Here is formula used that calculates new weights of query terms and produces optimal results by discarding non relevant terms. It is called **Rocchio Formula**.

*Aim:* - The aim of this formula is to increase weights of terms that occur in relevant documents and decrease the weights of terms occurring in non relevant documents.

*Equation*: -

$Q_a$ **(new) = x * $Q_a$ (old) + y * 1/ (RD) * ∑ $wt_{aRD}$ – z * 1/ (NRD) * ∑ $wt_{aNRD}$**

Where $Q_a$ (new) = New weight of query term a

$Q_a$ (old) = old weights of tern s

RD = Relevant documents judged by user

NRD = Non- Relevant documents judged by user

$wt_{aRD}$ = Weights of terms in relevant documents

$wt_{aNRD}$ = Weights of terms in non relevant documents

∑ $wt_{aRD}$ = All weights of RD are added together

∑ $wt_{aNRD}$ = All weights of NRD are added together

y = It is constant that gives average of weights of terms in RD

z = It is constant that gives average of weights of terms in NRD. The result is that we get negative weights and they will be discarded automatically.

## 3.6 Retrieval Module

It is module that retrieves final results/optimal queries that have been extracted after going through various phases. It ranks document according to similar queries and maintains index according to information symbols contained in that query.

### 3.6.1 Re-Use of Queries

**Need for Re-Use of Queries**: - The queries that were already expanded and refined according to user's requirements are optimized and stored anywhere. If user needs some information ion future, then what is way to retrieve those documents that satisfies query?

**Solution**: - Re-Use of queries.

**Analysis**: - The expanded and refined queries are stored in database that is called as Query database. The query base contains queries related to previously retrieved documents. These queries are called *Persistent Queries*.

**How to Use Persistent Queries with new Query?**

(a) If a new query is somewhat similar to persistent query, then result of new query is related to persistent query.

(b) If user new query is not similar to persistent query in any way, then system has to find persistent query from database that satisfies new query to some extent.

**How to check similar queries?**

*Using concept of Solution Region*: - When search for an optimal query begins, system retrieves number of queries instead of only one query. All those queries are described in query space. The region containing that query space is called Solution Region.

We can check similarity between queries as the new queries are compared with queries in solution region and if they get matched, then both queries are said to be similar.

## 4. EXPERIMENTAL ANALYSIS AND CALCULATIONS

Consider a given sets of data. We have to compute probabilities of relevant and non relevant terms and hence calculate weight function for each term.

Given data:

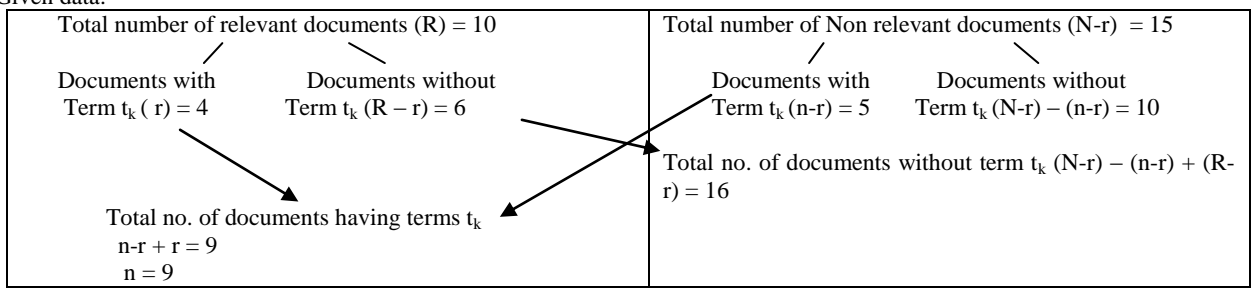

```
Total number of relevant documents (R) = 10        Total number of Non relevant documents (N-r) = 15
       /            \                                       /            \
Documents with   Documents without                  Documents with   Documents without
Term t_k ( r ) = 4   Term t_k (R – r) = 6           Term t_k (n-r) = 5   Term t_k (N-r) – (n-r) = 10

                                                   Total no. of documents without term t_k (N-r) – (n-r) + (R-r) = 16

        Total no. of documents having terms t_k
                 n-r + r = 9
                 n = 9
```

According to BI model,
Total number of documents N= 25
Total number of documents with term $t_k$ (n) = 9
Total number of relevant documents (R) = 10
Total number of relevant documents with term $t_k$ (r) = 4
From above data,
$P_k$ = Probability of term $t_k$ occurring in relevant documents
= 4/10 = 2/5

$U_k$ = Probability of term $t_k$ occurring in non - relevant documents
= 5/15 = 1/3
$X = P_k / (1 - P_k)$
= (2/5)/ (3/5) = 2/3
$Y = U_k / (1 - U_k)$
= (1/3) / (2/3) = ½
Odd Ratio or Weighting Function $W_k$ = X/Y = 4/3
Ranking function W = log (X/Y) = log (4/3) = 0.20068

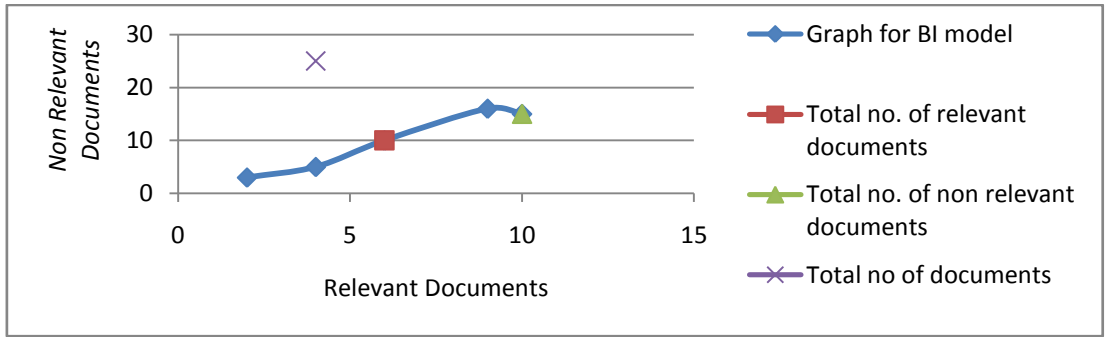

**Figure6: - Computation of Probabilities of terms graphically**

On the basis of above graph and probability values, we can find new weight function for terms from old weight function by using Rocchio Formula.

$Q_a$ **(new) = x * $Q_a$ (old) + y * 1/ (RD) * ∑ $wt_{aRD}$ – z * 1/ (NRD) * ∑ $wt_{aNRD}$**

Here $Q_a$ (old) = 4/3

Relevant Documents (RD) = 10

Non relevant documents (NRD) = 15

∑ $wt_{aRD}$ = 4 + 6 = 10

∑ $wt_{aNRD}$ = 5 + 10 = 15

X = 1, y = (4+6) / 2 = 5, z = (5+10) / 2 = 15/2 = 7.5

So, $Q_a$ (new) = 1 * (4/3) + 5 * (1/10) * 10 – 7.5 * (1/15) * 15

= 4/3 + 5 – 7.5 = - 3.7

Since new weight function is negative, so it is discarded and old function is considered as relevance function.

*Catchy Concept*: - The proposed High-Level Statistical Multimedia IR Model deals with the queries that have been expanded and refined according to user's requirements. In this way the queries can be reused. It is good idea if given queries is short. HOW THIS MODEL CAN BECOME SUITABLE FOR LONG QUERIES ALSO?

*Catchy Answer*: - The answer to above question is *Use of Random Variables*. These variables may be continuous as well as Discrete. The terms that are found in multimedia text documents can be treated as variables. If the terms are short or finite, then it is solved using concept of Discrete Random Variables. It simply means adding product of probabilities of various terms/queries used in document. In this way, expected value [E] of term function is calculated and we can determine its relevance.

For long queries, the concept of Continuous Random Variables can be used. Further, long queries may have some limit or they are infinite. For queries having limit, approximation is used. The terms are integrated to particular interval and produce results proximity to user's requirements.

For infinite long queries, various methods of calculating expected value [E] like Poisson distribution, Binomial distribution are employed.

In this way, both short queries as well as long queries can be reused and expanded.

## 5. CONCLUSION
The paper illustrates the working of proposed high level multimedia IR model consisting of various modules. Each module is described separately. This module provides extraction of relevant terms from huge collection of multimedia documents. Since multimedia documents produce information tokens that are different from text tokens, so those statistical approaches are shown in paper that analyses multimedia document and retrieves multimedia terms (text, images, and videos) from them.

The new model can replace ambiguities of traditional multimedia IR model that deals with information symbols only instead of maintaining relationships between them. It is beneficial in various aspects like there is module introduced in it for maintaining conceptual relationships between extracted terms and represents them using ontology. The model uses probabilistic approaches for calculating ranking of documents and retrieves optimal queries. The results are then presented to user.

## ABOUT THE AUTHORS

**Gagandeep Singh** has completed his B.Tech (CSE) from GTBIT affiliated to Guru Gobind Singh Indraprastha University, Delhi. His Research areas include Semantic Web, Information Retrieval, Data Mining, Remote Sensing (GIS) and Knowledge Engineering.

**Vishal Jain** has completed his M.Tech (CSE) from USIT, Guru Gobind Singh Indraprastha University, Delhi and doing PhD in Computer Science and Engineering Department, Lingaya's University, Faridabad. Presently, He is working as Assistant Professor in Bharati Vidyapeeth's Institute of Computer Applications and Management, (BVICAM), New Delhi. His research area includes Web Technology, Semantic Web and Information Retrieval. He is also associated with CSI, ISTE.